\documentclass{iau_FM}
\usepackage{graphicx}
\usepackage{tabulary}
\usepackage{natbib}

\bibpunct{(}{)}{;}{a}{,}{,}

\title[3D Simulations of Carbon Burning] %% give here short title %%
{The First 3D Simulations of Carbon Burning in a Massive Star}

\author[A. Cristini, C. Meakin, R. Hirschi, D. Arnett, C. Georgy, M. Viallet]   %% give here short author list %%
{A. Cristini$^{a,b\thanks{Email: a.j.cristini@ou.edu}}$, C. Meakin$^{c,d}$, R. Hirschi$^{a,e}$, D. Arnett$^d$, C. Georgy$^{f,a}$,\, M. Viallet$^g$}

\affiliation{$^a$\textsf{Astrophysics Group, Keele University, Lennard-Jones Laboratories, Keele, ST5 5BG, UK}\\
$^b$\textsf{Department of Physics and Astronomy, University of Oklahoma, Norman, OK 73019, USA}\\
$^c$\textsf{Karagozian \& Case, Inc., 700 N. Brand Blvd. Suite 700, Glendale, CA, 91203, USA}\\
$^d$\textsf{Department of Astronomy, University of Arizona, Tucson, AZ 85721, USA}\\
$^e$\textsf{Kavli IPMU (WPI), The University of Tokyo, Kashiwa, Chiba 277-8583, Japan}\\
$^f$\textsf{Geneva Observatory, University of Geneva, Maillettes 51, 1290 Versoix, Switzerland}\\
$^g$\textsf{Max-Planck-Institut f\"{u}r Astrophysik, Karl Schwarzschild Strasse 1, Garching, D-85741, Germany}}

\pubyear{2017}
\jname{The lives and death-throes of massive stars} 
\editors{J.J. Eldridge}
\begin{document}

\maketitle

\begin{abstract}
We present the first detailed three-dimensional hydrodynamic implicit large eddy simulations of turbulent convection for carbon burning. The simulations start with an initial radial
profile mapped from a carbon burning shell within a 15\,M$_\odot$ stellar evolution model. We considered 4 resolutions from 128$^3$ to 1024$^3$ zones. These simulations confirm that convective boundary mixing (CBM) occurs via turbulent entrainment as in the case of oxygen burning. 
The expansion of the boundary into the surrounding stable region and the entrainment rate are smaller at the bottom boundary because it is stiffer than the upper boundary. The results of this and similar studies call for improved CBM prescriptions in 1D stellar evolution models.
\end{abstract}

\renewcommand{\thefootnote}{\arabic{footnote}}
One-dimensional (1D) stellar evolution codes are currently the only way to simulate the entire lifespan of a star. This comes at the cost of having to replace complex, inherently three-dimensional (3D) processes, such as convection, rotation and magnetic activity, with generally simplified mean-field models. An essential question is ``how well do these 1D models represent reality?'' Answers can be found both in empirical and theoretical work. On the empirical front, we can investigate full star models, by comparing them to observations of stars under a range of conditions, in particular, asteroseismology. On the theoretical side, multi-dimensional simulations can be used to test 1D models under astrophysical conditions that are difficult to re-create in terrestrial laboratories. We present here the latter; three-dimensional hydrodynamic simulations of carbon shell burning in a 15\,M$_\odot$ star. 

The reasons for choosing carbon burning as opposed to other burning regions are: this phase of stellar evolution has never been studied before; cooling is dominated by neutrino losses, allowing radiative effects to be neglected (very high P\'{e}clet$^{\scriptsize \footnotemark}$ number); the initial composition and structure profiles of the shell are simpler than the more advanced stages as the composition in the region where the shell forms has been homogenised by the preceding convective helium burning core; and finally this shell plays an important role in setting the final mass of the iron core prior to the core-collapse event. Choosing a shell as opposed to a core burning region also affords the simulation of two distinct convective boundaries rather than one. 

\footnotetext{The P\'{e}clet number is the ratio of heat transfer through conduction to heat transfer through advective motions.}

We prepared the initial conditions by calculating a 15\,M$_\odot$, solar metallicity, non-rotating 1D stellar evolution model until the end of the oxygen burning phase using the Geneva stellar evolution code (GENEC; \citealt{2008Ap&SS.316...43E}). %Following a parameter study of various core and shell burning convective boundaries calculated in this stellar model, we decided that the second carbon burning shell would be the most interesting and worthwhile phase to study initially using detailed 3D hydrodynamic simulations. Choosing a shell as opposed to a core burning region affords the simulation of two distinct convective boundaries rather than one. 
The hydrodynamic simulations were calculated from the structure given by this stellar model during the growth of the carbon burning shell.
%following a remapping onto a finer mesh and interpolation onto a Eulerian grid. 
These simulations were calculated using the Prometheus MPI (PROMPI; \citealt{2007ApJ...667..448M}) code which solves the inviscid Euler equations using a finite-volume Eulerian solver which utilises the piecewise parabolic method of \citet{1984JCoPh..54..174C}. 
%The implicit large eddy simulation paradigm is utilised in order to simultaneously capture the integral scale and inertial range. The hydrodynamics solver is complemented by the Helmholtz equation of state \citep{1999ApJS..125..277T,2000ApJS..126..501T}. The nuclear energy generation rate due to $^{12}$C - $^{12}$C fusion is parameterised using a modified version of the formula given by \citet{1986nce..conf.....A,2009pfer.book.....M}. The composition is represented by three quantities: the average atomic mass, $\bar{A}$; average atomic number, $\bar{Z}$; and $^{12}$C mass fraction, but the change in $^{12}$C abundance due to nuclear burning is not accounted for as the advective time scale for mixing is much shorter than the nuclear burning timescale (small Damk\"{o}hler number). Cooling due to neutrino losses is parameterised according to the formula given by \citet{1967ApJ...150..979B}. Although the effects of rotation and magnetic fields are important in their own right and also due to their effect on convection, we do not include them in our models and focus on a purely hydrodynamic problem, this helps to reduce the computational complexity significantly.
We chose to model the domain within a plane-parallel, Cartesian geometry.
%with reflective vertical boundary conditions, periodic horizontal boundary conditions and a parameterised gravitational acceleration (inversely proportional to the radius). 
This `box-in-a-star' approach allows us to maximise the effective resolution at the convective boundaries, and ease the difficult Courant time scale at the inner boundary of the grid. More details on the PROMPI code and the model set-up can be found in \citet{2016arXiv161005173C}.
%A damping region is inserted along the bottom of the computational domain, here, velocities of propagating gravity mode waves are reduced in order to mimic their propagation out of the computational domain.

Simulations of turbulence involve some kind of initialisation of turbulent motions, followed by a transient phase whereby the global motion settles down into a quasi-steady state of turbulence. Initial test calculations of carbon burning revealed that the time-scale for this relaxation to the quasi-steady state was long, and therefore simulations of the quasi-steady state over time-scales that are statistically significant (several convective turnovers) would not be possible given our available computational resources. We therefore decided to boost the nuclear energy generation rate by a factor of 1000 in order to match that of oxygen burning; this reduces both the relaxation time and the convective turnover time. Such an artificial boost in luminosity does not affect the structure for the following reasons: hydrostatic equilibrium is still maintained; the entropy and composition profiles in the convective region remain flat; and the structure in the stable regions away from the boundary are determined by the evolutionary history of the model and are unaffected by the turbulence.

\begin{figure*}
\hspace{-10mm}
\includegraphics[width=0.32\textheight]{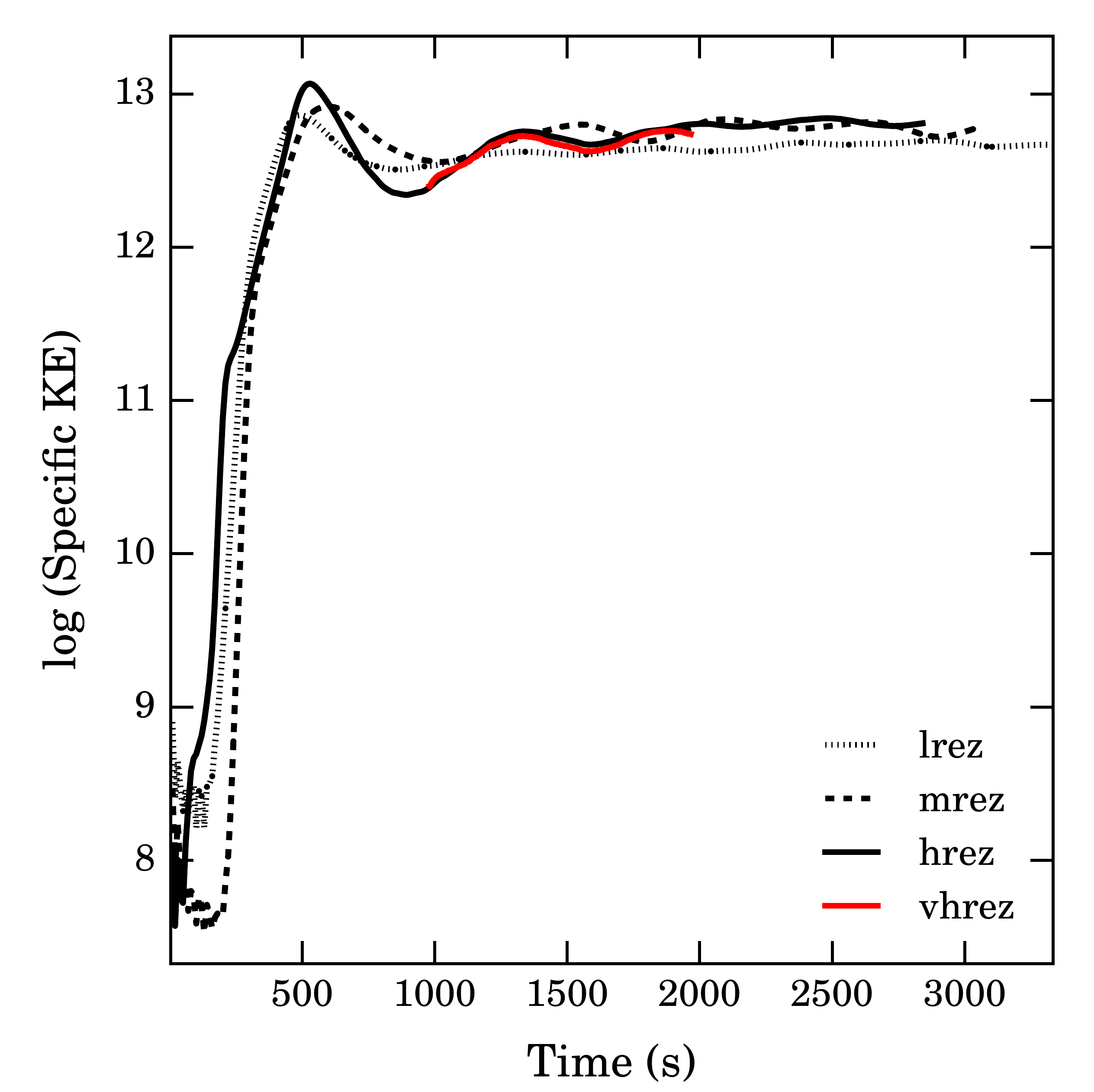}
\includegraphics[width=0.045\textheight,height=0.32\textheight]{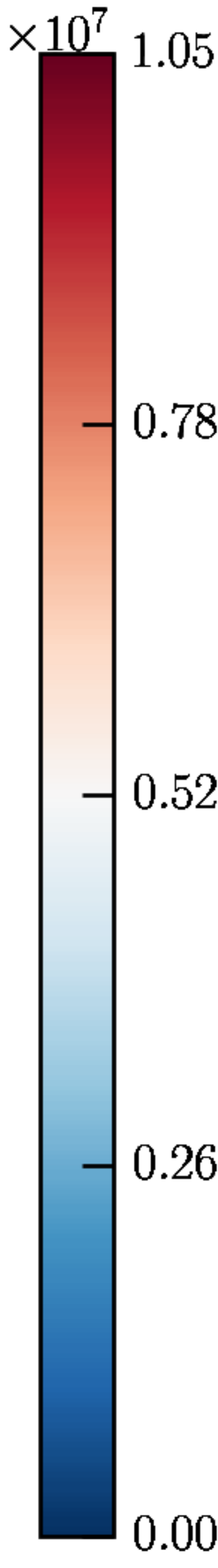}
\includegraphics[width=0.32\textheight,height=0.32\textheight]{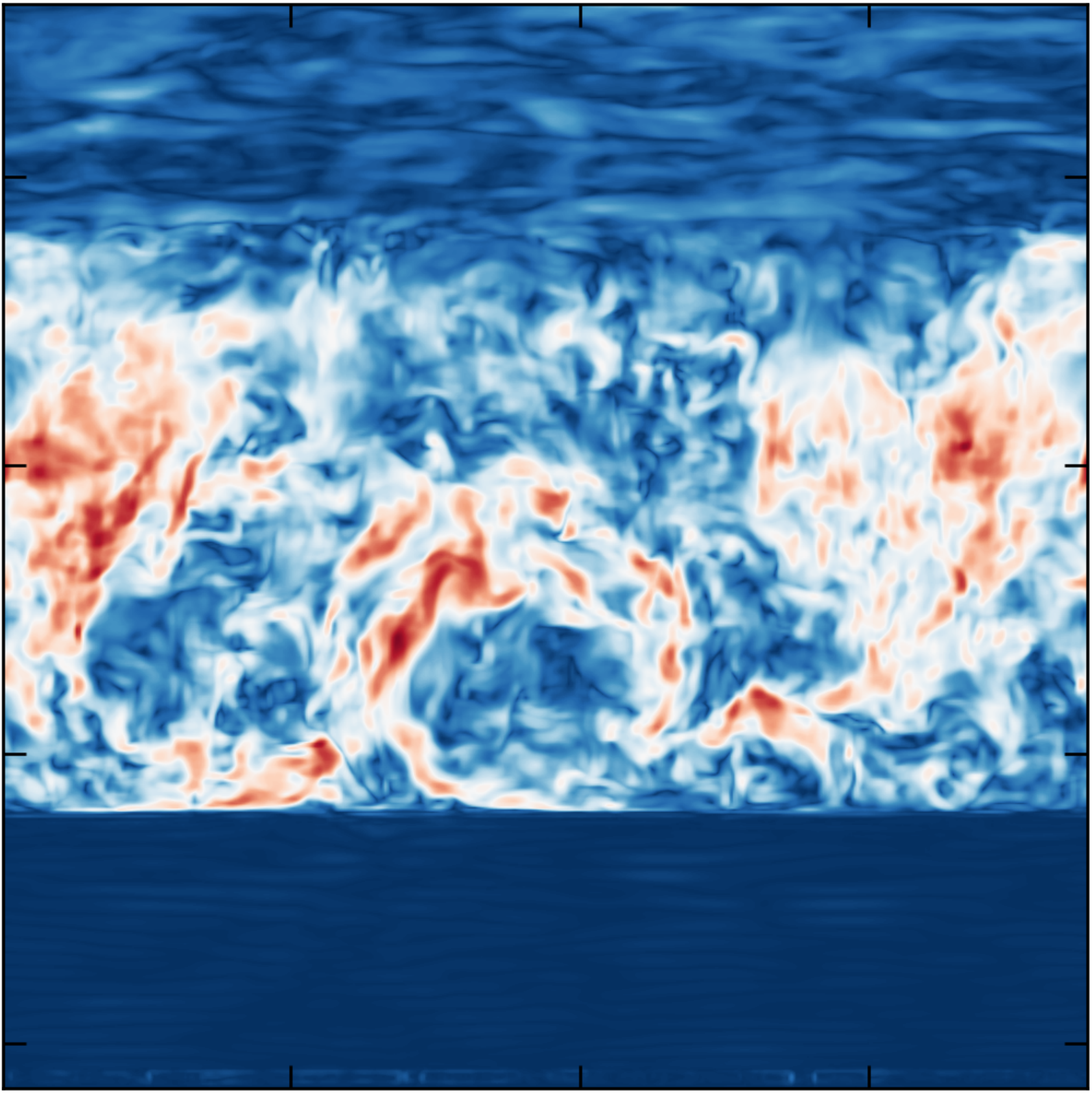} %,natwidth=277,natheight=277

\caption{\textit{Left:} Temporal evolution of the global specific kinetic energy: 
thin dashed - \textsf{lrez}; 
thick dashed - \textsf{mrez}; 
black solid - \textsf{hrez}; 
red solid - \textsf{vhrez}. The quasi-steady state begins after approximately 1,000$\,$s.
\textit{Right:} Vertical 
cross-section of the radial and horizontal component of the velocity vector field, $(v_r,v_y)$, at $t=2820$\,s. The colour-map represents the velocity magnitude in cm\,s$^{-1}$ \citep[taken from][]{2016arXiv161005173C}.}
\label{ek}
\end{figure*}

To test the dependence of our results on numerical resolution we simulated the carbon shell at four different resolutions. These models are named according to their resolution: \textsf{lrez} - 128$^3$, \textsf{mrez} - 256$^3$, \textsf{hrez} - 512$^3$ and \textsf{vhrez} - 1024$^3$. The temporal evolution of the global (integrated over the convective zone) specific kinetic energy for all of the models is presented in the left panel of Fig. \ref{ek}. The first $\sim$1000 seconds of evolution is characterised by the initial transient associated with the onset of convection. By $\sim$1250\,s, all of the models settle into the quasi-steady state of turbulence, characterised by semi-regular pulses in kinetic energy occurring on a time scale of the order of the convective turnover time. These pulses are associated with the formation and eventual breakup of semi-coherent, large-scale eddies or plumes that traverse a good fraction of the convection zone before dissipating. It is a phenomena that is typical of stellar convective flow \citep{2007ApJ...667..448M,2011ApJ...733...78A,2011ApJ...741...33A,2013ApJ...769....1V,2015ApJ...809...30A}. Although these simulations do not sample a large number of convective turnover times (between $\sim$2 and $\sim$6), resolution trends are still apparent. 

Some aspects of our models are sensitive to the grid resolution. At the lower convective boundaries of our models a spurious spike in dissipation appears at all resolutions (see figs. 7 and 8 in \citealt{2016arXiv161005173C}). This spike appears to be numerical and undermines the statistical analysis, although the general behaviour of the numerical dissipation is sane, and the discrepancy is localised. The spike reduces in amplitude and width with increasing resolution, suggesting convergence to a physically relevant solution. Our resolution study shows that a radial resolution of 512 zones is sufficient to resolve the upper boundary but a resolution of roughly 1500 zones is needed to fully resolve the lower boundary \citep[see][for details]{2016arXiv161005173C}. 

The qualitative description of convection and CBM is very different from that which describes the parameterisations that are used in stellar evolution models. The velocity magnitude, $\sqrt{v_r^2 + v_y^2}$ (where $v_r$ and $v_y$ are the radial and horizontal (in the y direction) velocities, respectively) of the \textsf{hrez} model is shown in the right panel of Fig. \ref{ek}.
Entrainment events (similar to those found for oxygen burning, see e.g. fig. 23 in \citealt{2007ApJ...667..448M}) can be seen in the convection zone (see e.g. bottom left of convective zone where material from below the convective zone is entrained upwards or top corners of the convective zones where the material is entrained from the top stable layer).  Strong flows can be seen in the centre of the convective region and shear flows can be seen over the entire convective region. These shear flows have the greatest impact at the convective boundaries, where composition and entropy are mixed between the convective and radiative regions. 

Turbulent entrainment at both boundaries pushes the boundary position over time into the surrounding stable regions. In order to calculate the boundary entrainment velocities, first the convective boundary positions must be determined in the simulations. In the 3D simulations, the boundary is a two-dimensional surface and is not spherically symmetric as in 1D stellar models. Thus the convective boundary position must be estimated.
In order to do this we first map out a two-dimensional horizontal boundary surface, $r_{j,k}=r(j,k)$, for $j=1,n_y$; $k=1,n_z$, where $n_y$ and $n_z$ are the number of grid points in the $y$ and $z$ directions. We estimate the radial position of the boundary at each horizontal coordinate coincides with the position where the average atomic weight, $\bar{A}$, is equal to the mean value of $\bar{A}$ between the convective and corresponding radiative zones. The boundary position at each time-step is then approximated as the horizontal mean, $\overline{r}_{j,k}$ (henceforth denoted as $\overline{r}$), over the boundary surface. We define the error in the estimated boundary position as the standard deviation ($\sigma$) from the horizontal surface mean, $\overline{r}$. 

%At the lower convective boundaries of our models a spurious spike in dissipation appears at all resolutions. This spike appears to be numerical and undermines the RANS analysis, although the general behaviour of the numerical dissipation
%is sane, and the discrepancy is localised. The spike reduces in amplitude and width with increasing resolution, suggesting convergence to a physically relevant solution. Our resolution study shows that a radial resolution of 512 zones is sufficient to resolve the upper boundary but a resolution of roughly 1500 zones is needed to fully resolve the lower boundary. 
%\begin{figure*}
%\caption{}
%\label{abar_cz}
%\end{figure*} %abar_movie_hrez.py

%The most prominent trend seen here is the kinetic energy peak associated with the initial transient, which increases as the grid is refined. This is not linked to the initial seed perturbations and is most likely related to the decreased numerical dissipation at finer zoning. A similar trend can also be seen in the quasi-steady turbulent state that follows the initial transient. Interestingly, in this case, a resolution dependence only appears to manifest for the lowest resolution model, \textsf{lrez}. This has an overall smaller amplitude of kinetic energy as well as a much smaller variance associated with the formation and destruction of large scale eddies. These properties can be naturally attributed to a higher numerical dissipation at a lower resolution.

 %reconstruct_RANS.py

The average atomic weight, $\bar{A}$, is used as an input variable in the Helmholtz equation of state \citep{1999ApJS..125..277T}, and so as the models evolve we calculate new boundary surface positions, $r_{j,k}$. Our method is a valid but not unique way in which to calculate the boundary positions \citep[e.g.][]{1998JAtS...55.3042S,2004JAtS...61..281F,2007ApJ...667..448M,2011PhRvE..84a6311L,2011JAtS...68.2395S,2011JPhCS.318d2061V,2014AMS...1935.JGJG,2015JFM...778..721G}.

\par
\begin{figure*}
\centering
\hspace{-10.45mm}\includegraphics[scale=0.275]{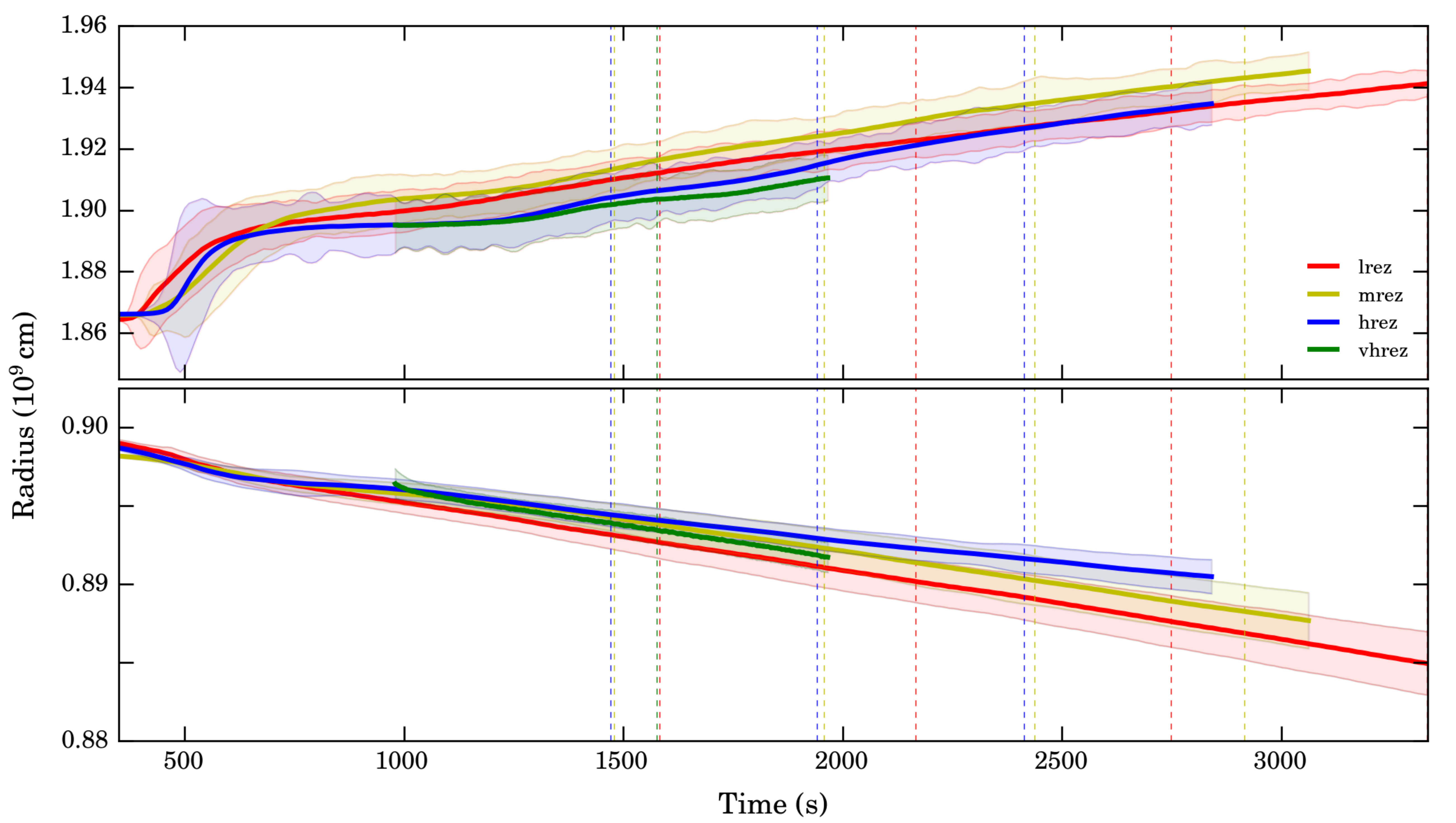}
\caption{Time evolution of the mean radial position of the upper convective boundary (top panel) and lower convective boundary (bottom panel), averaged over the 
horizontal plane for all four resolutions. Shaded envelopes are twice the standard 
deviation from the boundary mean. Vertical 
dotted lines indicate convective turnover times, taken from the beginning of the quasi-steady state at $\sim$1000s. The shaded areas represent 
the variation in the boundary height due to the fact that the boundary is not a flat surface. This can be compared 
to the ocean surface not being flat due to the presence of waves \citep[taken from][]{2016arXiv161005173C}.}
\label{rad}
\end{figure*}

 The variation in time of the average surface position, $\overline{r}$, of both boundaries is shown for all models in Fig. \ref{rad}. Positions are shown as solid lines and twice the standard deviation as the surrounding shaded envelopes. These small envelopes are due to the vertical extent of the boundary surface, which is not flat. Following the initial transient (\textgreater$\,1000\,\textrm{s}$) a quasi-steady expansion of the convective shell proceeds, convective turnovers are indicated by vertical dashed lines for each resolution in Fig. \ref{rad}. The shape of the boundary is further discussed in \citet{2016arXiv161005173C}.

 %buoy_n2.py

In conclusion, 1D stellar evolution models should include CBM at all convective boundaries, and turbulent entrainment should be accounted for in the advanced stages of massive star evolution. 
At the lower boundary of our models, which is stiffer, the entrainment is slower and the boundary width is narrower. This confirms the dependence of entrainment and mixing on the stiffness
of the boundary. Since the boundary stiffness will vary both in time and with the convective boundary considered, a single constant parameter is probably not going to correctly represent the dependence of the mixing on the instantaneous convective boundary properties. We suggest the use of the bulk Richardson number \citep{2016PhyS...91c4006C} in new prescriptions in order to include this dependence.\\
%\\\\\\\\\\\\\\

The authors acknowledge support from EU-FP7-ERC-2012-St Grant 306901. RH acknowledges support from the World Premier International Research Centre Initiative (WPI Initiative), MEXT, Japan. This work used 
the Extreme Science and Engineering Discovery Environment (XSEDE), which is supported by National Science Foundation grant number OCI-1053575. CM and WDA acknowledge support from NSF grant 1107445 at the 
University of Arizona. The authors acknowledge the Texas Advanced Computing Center (TACC) at The University of Texas at Austin (http://www.tacc.utexas.edu) for providing HPC resources that have contributed 
to the research results reported within this paper. MV acknowledges support from the European Research Council through grant ERC-AdG No. 341157-COCO2CASA. This work used the DiRAC Data Centric system at 
Durham University, operated by the Institute for Computational Cosmology on behalf of the STFC DiRAC HPC Facility (www.dirac.ac.uk). This equipment was funded by BIS National E-infrastructure capital grant 
ST/K00042X/1, STFC capital grants ST/H008519/1 and ST/K00087X/1, STFC DiRAC Operations grant ST/K003267/1 and Durham University. DiRAC is part of the National E-Infrastructure.

%### grid scale truncation errors are set as the numerical dissipation of the system, the errors near the boundary result in a spurious spike in dissipation that is greater than the dissipation in the convective region, suggesting that this implied dissipation is no longer mimicking the physical dissipation but is a result of poor resolution at the boundary.

\bibliography{references}

\end{document}